\documentclass[amsmath,amssymb,showpacs,showkeys,superscriptaddress,twocolumn]{revtex4}
\usepackage{epsfig}
\usepackage{bm}
\usepackage{amssymb}
\begin{document}
\title{Gamma-distribution and Wealth inequality}

\author{Anirban Chakraborti}
\email{achakraborti[at]yahoo.com}
\affiliation{Department of Physics, Banaras Hindu University, Varanasi-221 005, India}

\author{Marco Patriarca}
\email{marco.patriarca[at]gmail.com}
\affiliation{National Institute of Chemical Physics and Biophysics, R\"avala 10, 12618 Tallin, Estonia}

\begin{abstract}
We discuss the equivalence between kinetic wealth-exchange models, in which agents exchange wealth during trades, and mechanical models of particles, exchanging energy during collisions.
The universality of the underlying dynamics is shown both through a variational approach based on the minimization of the Boltzmann entropy
and a complementary microscopic analysis of the collision dynamics of molecules in a gas.
In various relevant cases the equilibrium distribution is the same for all these models, namely a $\gamma$-distribution with suitably defined temperature and number of dimensions.
This in turn allows one to quantify the inequalities observed in the wealth distributions and suggests that their origin should be traced back to very general underlying mechanisms: for instance, it follows that the smaller the fraction of the relevant quantity (e.g. wealth or energy) that agents can exchange during an interaction, the closer the corresponding equilibrium distribution is to a fair distribution.
\\
\\
Presented to the International Workshop and Conference on: Statistical Physics Approaches to Multi-disciplinary Problems, January 07 - 13, 2008, IIT Guwahati, India
\end{abstract}



\maketitle

\section{Introduction}

The most noted power law in economics is perhaps the Pareto law, first observed
by Vilfredo Pareto \cite{Pareto1897a,Pareto1897a_reprint,Pareto1971a_reprint} more than a century ago.
It was found that in an economy the higher end of
the distribution of wealth $f(x)$ follows a power-law
\begin{equation}
f(x)\sim x^{-1-\alpha },\label{eq:paretolaw}
\end{equation}
and $\alpha $ is an exponent (now known as the Pareto exponent) which Pareto
estimated to be $\approx 3/\,2$.
For the last hundred years the changes in value $\alpha \sim $ 3/\,2
in time and across the various capitalist economies seem to be small.
This implies that there is inequality in the wealth distribution, and only a few persons hold the
majority of wealth.

In 1931, Gibrat \cite{Gibrat1931a} suggested that while Pareto's law is valid only for the high wealth range, the middle wealth range is given by the probability density
\begin{equation}
  f(x)\sim \frac{1}{x \sqrt{2\pi \sigma^2}} \exp{ \left\{ -\frac{\log^2 (x/x_0)}{2\sigma^2}\right\} } ,\label{eq:gibratlaw}
\end{equation}
where $x_0$ is a mean value and $\sigma^2$ is the variance.
The factor $\beta=1/\sqrt{2\sigma^2}$ is also know an as the Gibrat index, and a small Gibrat index corresponds to a uneven wealth distribution.

An unequal wealth distribution is associated not only to these functions, but
also to any other one which is not of the form a Dirac $\delta$-function.
The problem of the appearance of inequalities seems therefore to be rather general and not necessarily related to a particular shape -- e.g. the power-law form of the Pareto law -- of the wealth distribution, despite distributions can vary from case to case assuming qualitatively different shapes.
In fact wealth distribution has always been a prime concern of economics.
Classical economists such as Adam Smith, Thomas Malthus and
David Ricardo were mainly concerned with factor wealth distribution, that is,
the distribution of wealth between the main factors of production, land, labor
and capital. Modern economists have also addressed this issue, but have been
more concerned with the distribution of wealth across individuals and
households. Important theoretical and policy concerns include the relationship
between wealth inequality and economic growth.

Wealth inequality metrics or wealth distribution metrics are techniques used
by economists to measure the distribution of wealth among the participants in
a particular economy, such as that of a specific country or of the world in
general. These techniques are typically categorized as either absolute
measures or relative measures, and in the literature one of the most important
debates is on the issue of measuring inequality. While one type deals with the
objective sense of inequality, usually employing some statistical measure of
relative variation of wealth, the other type deals with some indices that try
to measure inequality in terms of some normative notion of social welfare for
a given total of wealth.

\emph{Absolute criteria}.
Absolute measures define a minimum standard, and then calculate the number (or
percent) of individuals below this threshold. These methods are most useful
when determining the amount of poverty in a society. Examples include the
poverty line, which is a measure of the level of wealth necessary to subsist
in a society. It varies from place to place and from time to time, depending on
the cost of living and people's expectations.

\emph{Relative criteria}.
Relative measures compare the wealth of one individual (or group) with
the wealth of another individual (or group). These measures are most useful
when analyzing the scope and distribution of wealth inequality. Examples
include the Gini coefficient, which is a summary statistic used to quantify the
extent of wealth inequality depicted in a particular Lorenz curve. The Gini
coefficient is a number between 0 and 1, where 0 corresponds with perfect
equality (where everyone has the same wealth) and 1 corresponds with perfect
inequality (where one person has all the wealth, and everyone else has zero
wealth).

However, it has to be noted that ``wealth'' is here understood differently
respect to its common meaning: it represents the total amount of goods and
services that a person
receives, and thus there is not necessarily money or cash involved.
Services like public health and education are also counted in, and often
expenditure or consumption (which is the same in an economic sense) is used to
measure wealth. Thus, it is not clear how wealth should be defined. There is
also the question that ``Should the basic unit of measurement be households or
individuals?'' The Gini value for households is always lower than for
individuals because of wealth pooling and intra-family transfers. The metrics
will be biased either upward or downward depending on which unit of
measurement is used. These and many other criticisms need to be addressed for
the proper use of inequality measures in a well-explained and consistent way.

In the attempt to answer some such basic questions and provide a foundation to the complex issues related to the appearance of wealth inequalities,
various authors have independently formulated minimal models of wealth exchange \cite{Angle1983a,Angle1986a,Bennati1988a,Bennati1988b,Bennati1993a,Chakraborti2000a} which, while being general enough to catch some universal features of economic exchanges, are simple enough to be simulated numerically in detail and studied analytically.
In these models a set of agents $i=1,\dots,M$, representing individuals or companies whose state is defined by the respective wealth $x_i$, interact with each other  from time to time by exchanging (a part of) their wealths.
Such exchanges are defined by laws depending on the $\{x_i\}$ and also contain some random elements, as detailed below.
A striking analogy was recognized -- and actually motivated the introduction of some of these models -- between the statistical mechanics of molecule collisions in a gas and these minimal models of economy, which are therefore referred to sometimes as \emph{kinetic wealth-exchange models}.
Such an analogy, rather than for its peculiarity, should be noticed since it signals a possible universal statical mechanism in action in the dynamical evolution of systems composed by single units, from gas composed of molecules colliding with each other exchanging their energy to economic societies in which single units interact by exchanging wealth.
This analogy is being analysed here in more details than done previously \cite{Patriarca2004a}, and represents the goal of the investigations presented here.

We begin by recalling the main features of kinetic wealth-exchange models in Sec.~\ref{features}, concentrating on an example of model with a fixed \emph{saving propensity} $\lambda$.
In Sec.~\ref{variational} it is shown how the fact that for a saving propensity $\lambda > 0$ one obtains an equilibrium $\gamma$-distribution $\gamma_n(x)$ of order $n$, instead of the Boltzmann law $\sim \exp(-x/\langle x \rangle)$ obtained for $\lambda = 0$, actually strengthens the kinetic analogy between economy systems and a gas in $N(\lambda)$ dimensions, with $N(\lambda)$ a known function of $\lambda$ \cite{Patriarca2004a,Patriarca2004b}: through a general variational approach based on the Boltzmann entropy it is shown that the $\gamma$-distribution $\gamma_n(x)$ of order $n$ is the equilibrium canonical distribution of a system with a quadratic Hamiltonian $H(q_1,\dots,q_N)$ and $N = 2n$ degrees of freedom $(q_1,\dots,q_N)$.
The analogy is further discussed in Sec.~\ref{kinetic}, this time through a complementary microscopic approach based on the analysis of the dynamics of particle collisions in an $N$-dimensional space.
Through mechanical considerations only based on momentum and energy conservation, it is shown that collision dynamics in $N$ dimension can be recast in the form of the evolution laws of kinetic wealth-exchange models, both for $\lambda = 0$ and in the case with saving propensity $\lambda > 0$, corresponding to a number of effective dimensions $N(\lambda) > 2$.
Finally, in Sec.~\ref{conclusion}, conclusions are drawn.


\section{Main features of kinetic closed-economy models}
\label{features}
Simple social models of wealth exchange have been shown to well reproduce many features of real wealth distribution.
For instance, the exponential law $f(x) \sim \exp(-\beta x)$ observed at intermediate values of wealth is reproduced by a many-agent model system composed of $M$ agents, who are assumed to exchange wealth in pairs at each iteration, according to the wealth-conserving evolution equations~\cite{Dragulescu2000a}
\begin{eqnarray}  \label{sp1}
  x_i' &=& \epsilon (x_i + x_j) \, ,
  \nonumber \\
  x_j' &=& \bar{\epsilon} (x_i + x_j) \, .
\end{eqnarray}
Here $\epsilon \equiv 1 - \bar{\epsilon}$ is a uniform random number in $(0,1)$, $i$ and $j$ are the labels of two agents chosen randomly at each iteration, and $(x_i,x_j)$ and $(x_i',x_j')$ represent the corresponding wealths before and after a trade, respectively.

More general versions of this model assign a (same) saving propensity $\lambda > 0$ to all agents, which represents the minimum fraction of wealth saved during a trade~\cite{Angle1983a,Angle1986a,Angle1993a,Chakraborti2000a,Chakraborti2002a,Angle2002a}.
As an example, in the model of Ref.~\cite{Chakraborti2000a} the evolution law is
\begin{eqnarray}  \label{sp2}
  x_i' &=& \lambda x_i + \epsilon (1-\lambda) (x_i + x_j) \, ,
  \nonumber \\
  x_j' &=& \lambda x_j + \bar{\epsilon} (1-\lambda) (x_i + x_j) \, .
\end{eqnarray}
It is to be noticed that while the total wealth is still conserved during a trade, $x_i' + x_j' = x_i + x_j$, only a fraction $(1 - \lambda)$
of the initial total wealth is reshuffled between the two agents during the trade.
These models relax toward an equilibrium distribution well fitted by a $\gamma$-distribution $\gamma_n(x)$, as also noted by Angle~\cite{Angle1986a},
\begin{eqnarray}\label{f-global}
  \beta^{-1} f(x) &\equiv& \gamma_{n}(\xi)
  = \frac{1}{\Gamma(n)} \, \xi^{n-1} \exp( - \xi )
  \, ,
  \nonumber \\
  \xi &=& {\beta x} \, ,
\end{eqnarray}
where the scaling parameter is $\beta^{-1} = \langle x \rangle / n$ and the parameter $2n(\lambda) \!\equiv\! N(\lambda)$, as shown below, represents an effective dimension of the system and is explicitly given by~\cite{Patriarca2004a,Patriarca2004b}
\begin{eqnarray}
  n(\lambda) \equiv \frac{N(\lambda)}{2}
   = 1 + \frac{3 \lambda}{1 - \lambda} = \frac{1 + 2\lambda}{1 - \lambda} \, .
  \label{N}
\end{eqnarray}
As the saving propensity $\lambda$ varies in $\lambda \in [0,1)$, the effective dimension $N(\lambda)$ continuously assumes the values in the interval $N \in [1,\infty)$.

\begin{figure*}[ht]
\begin{center}
\includegraphics[angle=0,width=.3\textwidth]{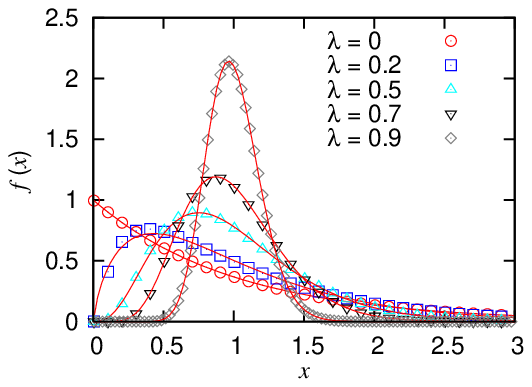}
\includegraphics[angle=0,width=.24\textwidth]{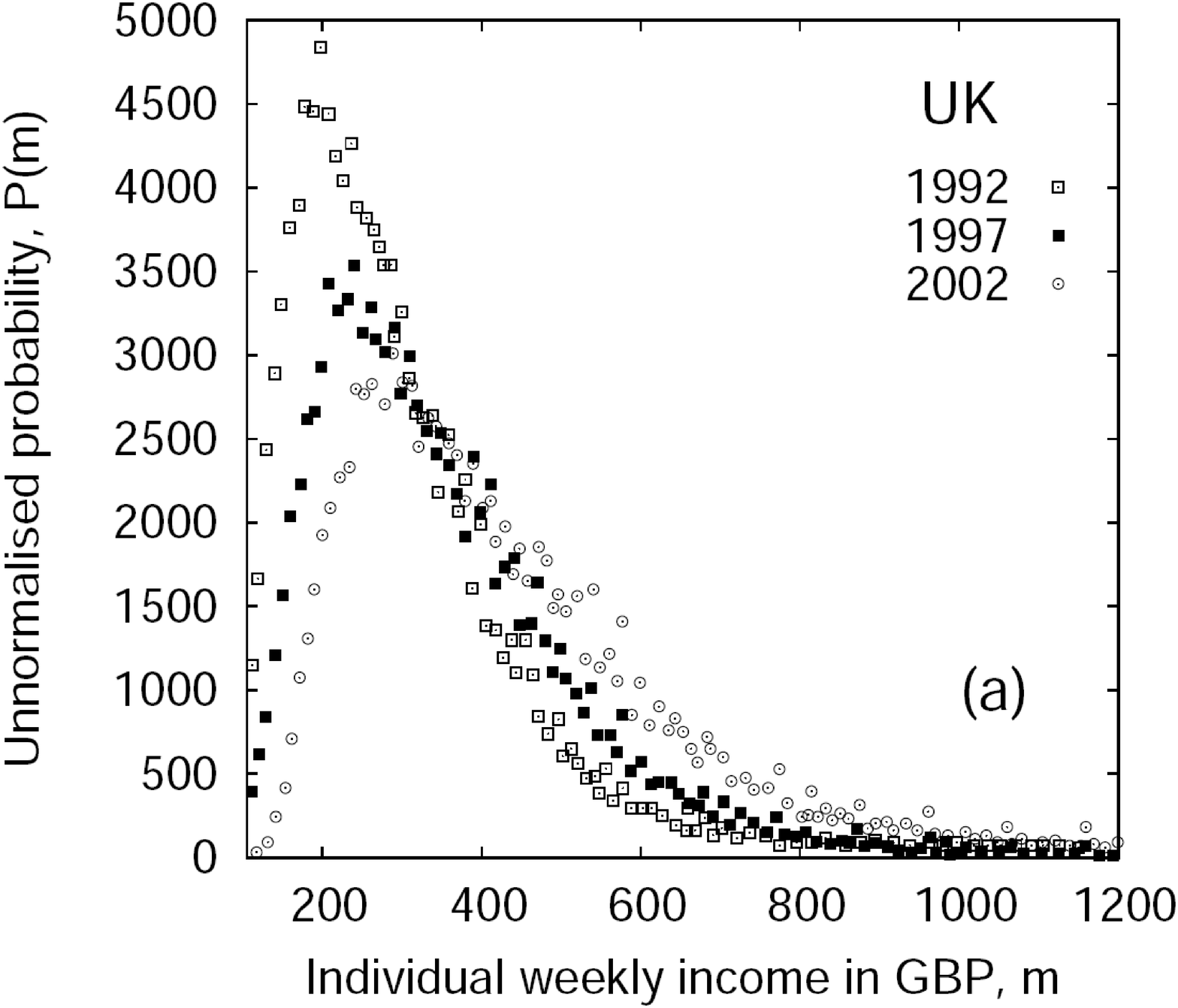}
\caption{(Left panel) Comparison between numerical results (various symbols) of the equilibrium wealth distribution in
the kinetic wealth-exchange models of closed economy and the corresponding
gamma functions (continuous lines) with suitable parameters (see text for details). (Right panel) Distribution $f(x)$ of individual weekly income or wealth $x$ in UK for
1992, 1997 and 2002; figure adapted from A. Chatterjee and B.K. Chakrabarti
cond-mat/0709.1543v2 using data adapted from G. Willis and J. Mimkes,
cond-mat/0406694.}
\label{realdata3}
\end{center}
\end{figure*}

A comparison between results of numerical simulation of wealth exchange models and the $\gamma$-distribution is shown in Fig.~\ref{realdata3}, together with
some real data for UK.
The the zero-saving propensity model of Eqs.~(\ref{sp1}) is described by the exponential curve for $\lambda = 0$.
The main difference of the curves with $\lambda > 0$ respect to the exponential distribution is the appearance of a peak: while the average wealth is unchanged (the system is closed and one still has $\langle x \rangle = {x_{\mathrm{tot}}} / M$, where $M$ is the total number of agents) the number of agents with a a wealth close to the average value increase or, in other words, the wealth distribution becomes more fair for larger $\lambda$'s and therefore larger values of the effective dimension $N$.
Eventually, as shown in Ref.~\cite{Patriarca2004a}, $f(x) \to \delta(x - \langle x \rangle)$ for $\lambda \to 1$ ($N \to \infty$) thus becoming a perfectly fair distribution.
As a consequence, measures of the inequality of the wealth distribution, such as the Gini coefficient, decrease for increasing $\lambda$ and tend to zero for $\lambda \to 1$.
In the rest of the paper we discuss the statistical mechanical interpretation of the equilibrium $\gamma$-distribution defined by Eqs.~(\ref{f-global}) and (\ref{N}).


\section{Mechanical analogy from the Boltzmann entropy}
\label{variational}
It may seem that in going from the exponential wealth distribution obtained for $\lambda = 0$ to the $\gamma$-distribution corresponding to a $\lambda > 0$ the link between wealth-exchange models and kinetic theory has been lost.
In fact, the $\gamma$-distribution $\gamma_{N/\,2}(x)$ represents but the Maxwell-Boltzmann equilibrium kinetic energy distribution for a gas in $N$ dimensions, as shown in the Appendix of Ref.~\cite{Patriarca2005a}.
Here a more general derivation of the $\gamma$-distribution $\gamma_{N/\,2}(x)$ is presented, in which it is shown, solely using a variational approach based on the Boltzmann entropy, that $\gamma_{N/\,2}(x)$ is the most general canonical equilibrium distribution of an $N$-dimensional system with a Hamiltonian quadratic in the coordinates.
Entropy-based variational approaches have been suggested (e.g. by Mimkes, Refs.~\cite{Mimkes2005a,Mimkes2005b}) to be relevant in the description and understanding of economic processes.
For instance, the exponential wealth distribution observed in real data and obtained theoretically in the framework of the Dragulescu-Yakovenko model discussed above, can also be derived through entropy considerations.
Considering a discrete set of $M$ economic subsystems which can be in one of $J$ possible different states labeled with $j = 1, 2, \dots, J$ and characterized by a wealth $x_j$, one can follow a line similar to the original Boltzmann's argumentation for the states of a mechanical system~\cite{Mimkes2005a}, by defining the total entropy as
\begin{equation} \label{W}
W\{m_j\} = \ln \frac{ M! } { m_1! m_2! \dots m_J! } \, ,
\end{equation}
where $m_j$ represents the occupation number of the $j$-th state; by variation of $W\{m_j\}$ respect to the generic $m_j$, with the constraints of conservation of the total number of systems $M = \sum_j m_j$ and wealth $x_{\rm tot} = \sum_j m_j x_j$, one obtains the canonical equilibrium distribution
\begin{equation} \label{m}
m_j \sim \exp(-\beta x_j) \, ,
\end{equation}
where $\beta = 1 /\langle x \rangle$ defines the temperature of the economic system.

Here we repeat the same argumentation for an ensemble of systems described by a continuous distribution $f(x)$, to show that this approach not only can reproduce the exponential distribution, but also the $\gamma$-distribution obtained in the framework of wealth-exchange models with a saving parameter $\lambda > 0$, a natural effective dimension $N > 2$ being associated to systems with $\lambda > 0$.

The representative system is assumed to have $N$ degrees of freedom $(q_1,\dots,q_N)$ and a homogeneous quadratic Hamiltonian, that for convenience is written in the rescaled form
\begin{equation} \label{X}
  x = X(q^2) \propto \frac{1}{2} q^2  = \frac{1}{2} (q_1^2+\dots+q_N^2) \equiv \frac{1}{2} \, q^2 \, ,
\end{equation}
where $q = (q_1^2 + \dots + q_N^2)^{1/\,2}$ is the distance from the origin in the $N$-dimensional $q$-space.
As a mechanical example, one can think of the $N$ momentum Cartesian components $(p_1,\dots,p_N)$ and the corresponding kinetic energy function $K = (p_1^2+\dots+p_N^2)/\,2m$, where $m$ is the particle mass;
or of the Cartesian coordinates $(q_1,\dots,q_N)$ of an isotropic harmonic oscillator with elastic constant $\kappa$ and potential energy $U = \kappa (q_1^2+\dots+q_N^2)/\,2$.
It can be checked -- e.g. using the Stirling approximation for the factorial function -- that in the limit of large occupation numbers, the discrete version (\ref{W}) of the Boltzmann distribution becomes proportional to $-\int dy f(y) \ln f(y)$, where the continuous variable $y$ replaces the discrete label $j$.
For an $N$-dimensional system, the entropy will be proportional to $-\int dq_1\dots\int dq_N f_N(q_1,\dots,q_N) \ln[f_N(q_1,\dots,q_N)]$, where $f_N(q_1,\dots,q_N)$ is the probability distribution in the $N$-dimensional space.
Then the constraints on the conservation of the total number of systems reads $\int dq_1\dots\int dq_N f_N(q_1,\dots,q_N) =$~const and that on total wealth is $\int dq_1\dots\int dq_N f_N(q_1,\dots,q_N) X(q^2) = x_{\rm tot}$.
In the end, using the Lagrange method, the equilibrium distribution density can be derived by functional variation respect to $f_N(q_1,\dots,q_N)$ of the effective functional
\begin{eqnarray} \label{Sn}
  S_\mathrm{eff}[f_N] =
  \int dq_1\dots\int dq_N f_N(q_1,\dots,q_N)\times
  \nonumber \\
  \times\{ \ln[f_N(q_1,\dots,q_N)] + \mu + \beta X(q) \} \, ,
\end{eqnarray}
where $\mu$ and $\beta$ are two Lagrange multipliers.
Before carrying out explicitly the variation it is convenient to simplify the problem exploiting the invariance of the Hamiltonian, only depending on the variable $q$, under rotations in the $q$-space.
The equilibrium probability density is expected to depend solely on the distance $q$ as well, i.e., $f_N(q_1,\dots,q_N) = f_N(q)$.
This allows the reformulation of the variational principle as an effective one-dimensional problem in the variable $q$, by transforming Cartesian to polar coordinates, so that
$$
\int dq_1\dots\int dq_N \, (\dots) \to \int_{0}^{+\infty} dq \int d\Omega \, (\dots) \, ,
$$
where the integration in $\Omega$ is over the total solid angle in the $N$-dimensional space.
After integration of the $N-1$ angular variables, one is left with
\begin{eqnarray} \label{func4}
  S_\mathrm{eff}[f_1] = \! \int_{0}^{+\infty} \!\!\!\! dq \, f_1(q)
  \left[ \! \ln\left( \frac{f_1(q)}{\sigma(N) q^{N-1}} \right) \!+\! \mu \!+\! \beta X(q) \! \right] \,
\end{eqnarray}
where $\sigma(N) = \int d\Omega = 2\pi^{N/\,2} / \, \Gamma(N/\,2)$ represents the surface of a hyper-sphere in $N$ dimensions~\cite{Hypersphere} with unit radius and we have expressed the probability density $f_N(q_1,\dots,q_N) = f_N(q)$ in the $N$-dimensional space in terms of the reduced probability density $f_1(q)$ for the variable $q$,
$$
f_1(q) \equiv \int d\Omega \, f_N(q) = \sigma(N) \, q^{N-1} f_N(q) \, .
$$
Finally, moving from $q$ to the energy variable $x = X(q^2) = q^2/\,2$, the corresponding probability density is
$$
f(x)
= \frac{dq(x)}{dx} \left.f_1(q)\right|_{q = q(x)}
= \frac{\left.f_1(q)\right|_{q = q(x)}}{\sqrt{2x}} \, ,
$$
where $q(x)$ is obtained by inverting Eq.~(\ref{X}).
In terms of the new variable $x$ and distribution $f(x)$, the effective functional (\ref{func4}) becomes
\begin{eqnarray} \label{func6b}
  S_\mathrm{eff}[f] = \! \int_{0}^{+\infty} \!\!\!\!\! dx \, f(x)
  \left[\! \ln\left( \frac{f(x)}{\sigma(N) x^{N/\,2-1}} \right)
  \!+\! \mu \!+\! \beta x \right] \!,
\end{eqnarray}
Variation of this functional,  $\delta S_\mathrm{eff}[f]/\delta f(x) = 0$, leads to the equilibrium $\gamma$-distribution (\ref{f-global}) with dimensionless variable $\xi=\beta x$ and index $n = N/\,2$.
As a simple example of application of this formula, one can obtain the Maxwell-Boltzmann probability density in three dimensions $f_3(K)$ for the kinetic energy $K$ letting $N = 3$.
In turn this suggests the interpretation of the parameter $N(\lambda)$ of wealth-exchange models as an effective dimension of the system and of the Lagrange multiplier $\beta^{-1}$ as the effective temperature, as it is in fact consistently recovered according to the equipartition theorem,
\begin{eqnarray}
  \beta^{-1} \equiv T =\frac{2 \langle x \rangle}{N}
   \, .
  \label{T}
\end{eqnarray}
%


\section{Mechanical analogy from collisions in $N$ dimensions}
\label{kinetic}
The deep analogy between kinetic wealth-exchange models of closed economy systems, where agents exchange wealth at each trade, and kinetic gas models,  in which energy exchanges take place at each particle collisions, can be further investigated and justified by studying the microscopic dynamics of interacting particles.
In this section we make more rigorous an argumentation only mentioned in Ref.~\cite{Patriarca2004a}.

In one dimension, particles undergo head-on collisions, in which they can exchange the total amount of energy they have, i.e. a fraction $\omega = 1$ of it.
Alternatively, one can say that the minimum fraction of energy that a particle saves in a collision is in this case $\lambda \equiv 1 - \omega = 0$.
In the framework of wealth-exchange models, this case corresponds to the model of Dragulescu and Yakovenko mentioned above~\cite{Dragulescu2000a}, in which the \emph{total} wealth of the two agents is reshuffled during a trade.

In an arbitrary (larger) number of dimensions, however, this does not take place, unless the two particles are travelling exactly along the same line in opposite verses.
On average, only a fraction $\omega = (1-\lambda) < 1$ of the total energy will be lost or gained by a particle during a collision,
that is most of the collisions will be practically characterized by an energy saving parameter $\lambda > 0$.
This corresponds to the model of Chakraborti and Chakrabarti~\cite{Chakraborti2000a},
in which there is a fixed maximum fraction $(1-\lambda) > 0$ of wealth which can be reshuffled.

Consider a collision between two particles in an $N$-dimensional space, with initial velocities represented by the vectors ${\bf v}_{(1)} = (v_{(1)1}, \dots, v_{(1)N})$ and ${\bf v}_{(2)} = (v_{(2)1}, \dots, v_{(2)N})$.
For the sake of simplicity, the masses of the two particles are assumed to be equal to each other and will be set equal to 1, so that momentum conservation implies that
\begin{eqnarray}  \label{v1}
  {\bf v}_{(1)}' &=& {\bf v}_{(1)} + \Delta {\bf v} \, ,
  \nonumber \\
  {\bf v}_{(2)}' &=& {\bf v}_{(2)} - \Delta {\bf v} \, ,
\end{eqnarray}
where ${\bf v}_{(1)}'$ and ${\bf v}_{(2)}'$ are the velocities after the collisions and $\Delta {\bf v}$ is the momentum transferred.
Conservation of energy implies that ${\bf v}_{(1)}'^{\,2} + {\bf v}_{(2)}'^{\,2} = {\bf v}_{(1)}^2 + {\bf v}_{(2)}^2$ which, by using Eq.~(\ref{v1}), leads to
\begin{eqnarray}  \label{v2}
  \Delta {\bf v}^2 + ({\bf v}_{(1)} - {\bf v}_{(2)}) \cdot \Delta {\bf v} = 0 \, .
\end{eqnarray}
Introducing the cosines $r_i$ of the angles $\alpha_i$ between the momentum transferred $\Delta {\bf v}$ and the initial velocity ${\bf v}_{(i)}$ of the $i$-th particle ($i = 1,2$),
\begin{eqnarray}  \label{cos}
  r_i = \cos\alpha_i  = \frac{{\bf v}_{(i)} \cdot \Delta {\bf v}}{v_{(i)} \, \Delta v} \, ,
\end{eqnarray}
where $v_{(i)} = |{\bf v}_{(i)}|$ and  $\Delta v =  |\Delta {\bf v}|$, and using Eq.~(\ref{v2}), one obtains that the modulus of momentum transferred is
\begin{eqnarray}  \label{v3}
  \Delta v = - r_1 v_{(1)} + r_2 v_{(2)} \, .
\end{eqnarray}
From this expression one can now compute explicitly the differences in particle energies $x_i$ due to a collision,
that are the quantities $x_i' - x_i \equiv ({\bf v}_{(i)}'^{\,2} - {\bf v}_{(i)}^2)/\,2$.
With the help of the relation (\ref{v2}) one obtains
\begin{eqnarray}  \label{x1}
  x_1' &=& x_1 + r_2^2 \, x_2 - r_1^2 \, x_1 \, ,
  \nonumber \\
  x_2' &=& x_2 - r_2^2 \, x_2 + r_1^2 \, x_1 \, .
\end{eqnarray}
Comparison with Eqs.~(\ref{sp1}) for the kinetic model of Dragulescu and Yakovenko clearly shows their equivalence -- consider that also here the $r_i$'s are in the interval $(0,1)$ and, furthermore, they can be considered as random variables, if a hypothesis of molecular chaos is made concerning the random initial directions of the two particles entering the collision.

However, the $r_i$'s are not uniformly distributed and their most probable value drastically depends on the space dimension: the greater the dimension $N$, the more unlikely it becomes that the corresponding values $\langle r_i \rangle$ assume values close to 1 and the more probable that instead they assume a small value close to $1/N$.
This can be seen by computing their average -- over the incoming directions of the two particles or, equivalently, on the orientation of the initial velocity ${\bf v}_{(i)}$ of one of the two particles and of the momentum transferred $\Delta {\bf v}$.
In $N$ dimensions, the Cartesian components of a generic velocity vector ${\bf v} = (v_1, v_2, \dots, v_N)$ are related to the corresponding hyper-spherical coordinates -- the velocity modulus $v$ and the $(N-1)$ angular variables $\varphi_j$ -- through the following relations,
\begin{eqnarray}  \label{v4}
  v_1 &=& v \cos\varphi_1 \, ,
  \nonumber\\
  v_2 &=& v \sin\varphi_1 \cos\varphi_2\, ,
  \nonumber\\
  &\dots&
  \nonumber\\
  v_N &=& v \sin\varphi_1 \sin\varphi_2 \dots \cos\varphi_N\, .
\end{eqnarray}
Using these transformations to express the initial velocity ${\bf v}_{(1)}$ of the first particle and the momentum transferred $\Delta {\bf v}$ in terms of their respective moduli $v_{(1)}$ and $\Delta v$ and angular variables $\{\phi_i\}$ and $\{\theta_i\}$, the expression (\ref{cos}) for the cosine $r_1$ becomes
\begin{eqnarray}  \label{v5}
  &&\!\!\!\!r_1 = \cos\alpha_1
  \nonumber \\
  &&\!\!\!\!=\! \cos\phi_1 \cos\theta_1 +
  \nonumber\\
  &&\!\!\!\!=\! [\sin\phi_1 \cos\phi_2 ] [\sin\theta_1 \cos\theta_2 ] +
  \nonumber\\
  &&\!\!\!\!\dots
  \nonumber\\
  &&\!\!\!\!=\! [\sin\phi_1 \sin\phi_2 \dots \cos\phi_N] [\sin\theta_1 \sin\theta_2 \dots \cos\theta_N] .
\end{eqnarray}
The average of the square cosine $r_1^2$ is performed by first taking the square of (\ref{v5}), integrating over the angular variables, considering that the $N$-dimensional volume element is given by
$$
d^{N}v = v^{N-1} dv \prod_{j=1}^{N-1} [\sin\varphi_j]^{{j-1}} d\varphi_{j} \, ,
$$
and finally by dividing by the total solid angle.
In the integration only the squared terms survive, obtained from the square of (\ref{v5}), since all the cross-terms are zero after integration --  they contain at least one integral of a term of the form $\sin\varphi\cos\varphi = \sin(2\varphi)/\,2$ which averages to zero.
By direct integration over the angle $\phi_i$ and $\theta_i$, it can be shown that in $N$ dimensions
$$
\langle r_1^2 \rangle = \langle \cos^2\alpha_1 \rangle = \frac{1}{N} \, .
$$
This means that the center of mass of the distribution for $r_1$, considered as a random variable due to the random initial directions of the two particles, shifts toward smaller and smaller values as $N$ increases.
The $1/N$ dependence of $\langle r_1^2 \rangle$ can be compared with the wealth-exchange model with $\lambda > 0$.
There a similar relation is found between the average value of the corresponding coefficients $\epsilon (1 - \lambda)$ or  $\bar\epsilon (1 - \lambda)$ in the evolution equations (\ref{sp2}) for the wealth exchange and the effective dimensions $N(\lambda)$: since $\epsilon$ is a uniform random number in $(0,1)$, then $\langle \epsilon \rangle = 1/\,2$ and from Eq.~(\ref{N}) one finds $\langle \epsilon (1 - \lambda) \rangle = (1-\lambda)/\,2 = 6/(N+4)$.


\section{Conclusion}
\label{conclusion}

The appearance of wealth inequalities in the minimal models and examples of closed economy systems considered above appears to reflect a general statistical mechanism taking place for a wide class of stochastic exchange law -- besides closed economy models -- in which the state of the $M$ units $\{i\}$ is characterized by the values $\{x_i\}$ of a certain quantity $x$ (e.g. wealth or energy) exchanged when units interact with each other.
The mechanism involved seems to be quite general and leads to equilibrium distributions $f(x)$ with a broad shape.
In the special but important case of systems with a homogeneous quadratic Hamiltonian -- or equivalently with evolution laws linear in the quantities $x_i$ -- and $N$ (effective) degrees of freedom, the canonical equilibrium distribution is a $\gamma$-distribution $\gamma_n(x)$ of order $n = N/2$.
The corresponding distribution for the closed economy model with a fixed saving propensity $\lambda$ has the special property that it becomes a fair (Dirac-$\delta$) distribution when $\lambda \to 1$ or $N(\lambda) \to \infty$.
The possibility for single units to exchange only a fraction of their wealth during a trade -- corresponding from a technical point of view to a wealth dynamics in a space with larger effective dimensions $N$ -- seems to be the key element which makes the wealth distribution less inequal.


\end{document}